\documentclass[aps,prl,twocolumn,showkeys,floatfix,nofootinbib]{revtex4}

\usepackage{graphicx}
\usepackage{dcolumn}
\usepackage{bm}

\newcommand{\kB}{k_{\text{B}}}
\newcommand{\rWS}{r_{\text{WS}}}

\begin{document}


\title{Atom-in-jellium equations of state and melt curves in the white dwarf regime}

\date{September 12, 2019; revisions to March 4, 2021 
   -- LLNL-JRNL-791167}

\author{Damian C. Swift}
\email{dswift@llnl.gov}
\author{Thomas Lockard}
\author{Sebastien Hamel}
\author{Christine J. Wu}
\author{Lorin X. Benedict}
\author{Philip A. Sterne}
\author{Heather D.~Whitley}
\affiliation{%
   Lawrence Livermore National Laboratory,
   7000 East Avenue, Livermore, California 94551, USA
}

\begin{abstract}
Atom-in-jellium calculations of the electron states,
and perturbative calculations of the Einstein frequency, were used to
construct equations of state (EOS) from around $10^{-5}$ to $10^7$\,g/cm$^3$
and $10^{-4}$ to $10^{6}$\,eV for elements relevant to white dwarf (WD) stars.
This is the widest range reported for self-consistent electronic shell structure calculations.
Elements of the same ratio of atomic weight to atomic number were predicted
to asymptote to the same $T=0$ isotherm, suggesting that, contrary to
recent studies of the crystallization of WDs, the amount of gravitational energy
that could be released by separation of oxygen and carbon is small.
A generalized Lindemann criterion based on the amplitude of
the ion-thermal oscillations calculated using atom-in-jellium theory,
previously used to extrapolate melt curves for metals,
was found to reproduce previous thermodynamic studies of the melt curve
of the one component plasma with a choice of vibration amplitude
consistent with low pressure results.
For elements for which low pressure melting satisfies the same amplitude criterion,
such as Al, this melt model thus gives a likely estimate
of the melt curve over the full range of normal electronic matter;
for the other elements, it provides a useful constraint on the melt locus.
\end{abstract}


\maketitle

\section{Introduction}
Interpretations of the populations and characteristics of stars rely on
simulations of the structure and evolution of individual, representative stars.
These simulations involve models of several types of physical behavior,
including the rate of different thermonuclear reactions, the rate of thermal
transport as dominated by the photon opacity or convection (depending on
the viscosity), and the equation of state (EOS) relating the pressure to the
mass density, temperature, and composition.

In the center of most stars and remnants such as white dwarfs (WD),
atoms are expected to be completely ionized, and the EOS should then be 
well understood \cite{hoteos}.
However, nearer the surface, atoms are partially-ionized, and the EOS
is more complicated.
Adequate models of the EOS in this regime are important in order to understand
the cooling rate of WDs and hence the population distribution.
The EOS is also important to understand proposed mechanisms of supernova
formation by accretion of matter onto a WD \cite{Foley2012}.

Considerable progress has been made in the last few years on rigorous methods
of calculating the EOS at individual states of density, temperature, and
composition.
The current state-of-the-art techniques include
path integral Monte Carlo (PIMC) and quantum molecular dynamics (QMD).
In PIMC, the Fermionic nature of the many-electron
density matrix is treated explicitly \cite{pimc}.
QMD has been used
extensively to simulate thermodynamic states, generally employing
density functional theory (DFT) 
with pseudopotentials to subsume the inner electrons and thus reduce the
number represented explicitly and hence the computational effort \cite{qmd}.
Nevertheless, such simulations can often involve $o(10^{16})$ floating point
operations per state, increasing with the number of electrons represented
explicitly.
It is not currently practicable to construct EOS directly with these 
techniques over ranges wide enough and tabulations fine enough to simulate
stars.

In practice, the
EOS used in star simulations use different physical methods and
approximations in different regimes.
A difficulty with this approach is that different techniques typically do not
connect seamlessly, and the resulting discrepancies
(usually manifesting as discontinuities in derivatives of the EOS)
can cause numerical problems in simulations.

Recent PIMC and QMD results have indicated that the simpler approach of
calculating the electron states for a single atom in a spherical cavity
within a uniform charge density of ions and electrons, representing the
surrounding atoms, reproduces the electronic component of 
their more rigorous EOS \cite{Benedict2014,Driver2017}.
This atom-in-jellium approach \cite{Liberman1979} can be extended to
estimate the ion-thermal energy from the Hellmann-Feynman force on the ion
when perturbed from the center of the hole in the jellium, which gives
an Einstein temperature.
We have previously extended this approach to predict the asymptotic freedom
of the ions at high temperature from the amplitude of the oscillations,
and thus the decrease in the ionic heat capacity from 3 to $\frac 32$\,$\kB$\ per
atom \cite{Swift_cowanx_2019} and in this way have constructed wide-range
EOS in agreement with PIMC and QMD results where available \cite{Swift_ajeos_2019}, for $o(10^9)$ floating-point operations per state.

The atom-in-jellium calculations employ DFT, and are thus inherently less
rigorous than PIMC, but they are relativistic and all-electron, and 
therefore in that respect are
more complete in their representation of the electrons than 
most QMD calculations, which employ pseudopotentials and solve the
non-relativistic Kohn-Sham equations.
We have previously constructed atom-in-jellium EOS tabulated with a range
and density similar to that of general purpose EOS in the {\sc sesame}
library \cite{sesame}:
$10^{-4}$ to $10^3$ times the ambient, condensed density with twenty points
per decade,
and temperatures of $10^{-3}$ to $10^5$\,eV with ten points per decade.
These EOS included carbon \cite{Swift_ajeos_2019}
and oxygen \cite{Lockard_cryoeos_2019}, 
which are both important in WDs.

\section{Wider-range EOS prediction}
WDs are thought to reach a density $o(10^7)$\,g/cm$^3$ in the center,
which is four orders of magnitude higher than our previous study and
than is typically the limit for general-purpose EOS \cite{sesame,leos},
and a temperature $o(10^6)$\,eV, an order of magnitude higher.

We found previously that the atom-in-jellium calculations ran more robustly
when constructing each isochore from high temperature to low.
Starting at $10^6$\,eV instead of $10^5$ further improved the robustness,
and we were able to extend the previous EOS to near $10^7$\,g/cm$^3$ and also
to lower densities, before encountering numerical difficulties.
The calculations were also continued an extra decade down in temperature,
to $10^{-4}$\,eV instead of $10^{-3}$, or 1.15\,K, 
to investigate the robustness of the calculation and avoid any need to
extrapolate down to the cosmic background temperature.

\begin{figure}
\begin{center}\includegraphics[scale=0.72]{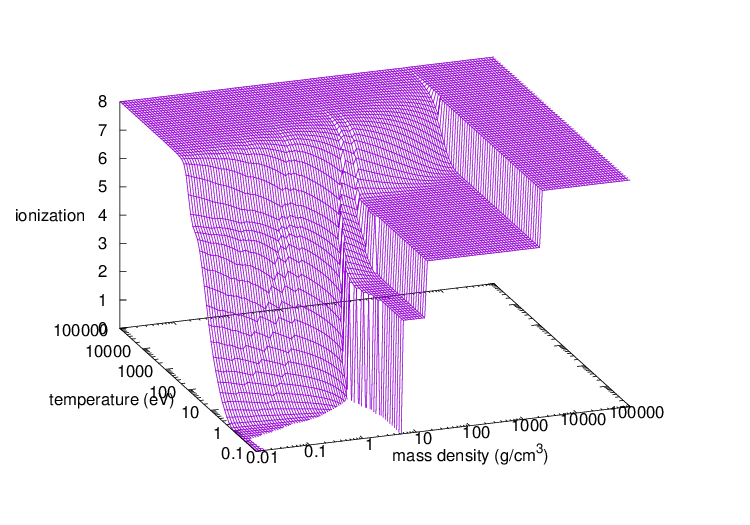}\end{center}
\caption{Calculated ionization of O as a function of mass density and
   temperature.
   Complete ionization is the region where the one-component plasma
   treatment is applicable.}
\label{fig:ozeff}
\end{figure}

EOS were constructed in this way for H, He, C, O, Ne, and Mg,
which are all significant constituents of WDs.
The EOS were entirely
consistent in their construction, and thus free of discontinuities from
the patching together of different techniques in different regimes.
For example, at low temperatures the ionization is predicted to change 
abruptly with compression,\footnote{%
Abrupt changes in ionization are not manifested as discontinuities
in pressure or heat capacity,
because the properties of a high-energy bound state
are similar to those of a low-energy free state:
the wavefunction of the bound state extends significantly far from the ion,
and the wavefunction of the free state is affected significantly by the
potential of the ion.
}
but varies more smoothly at higher temperatures (Fig.~\ref{fig:ozeff}).
These different behaviors arise naturally within this single, wide-ranging
model of electronic structure.

\begin{figure}
\begin{center}\includegraphics[scale=0.72]{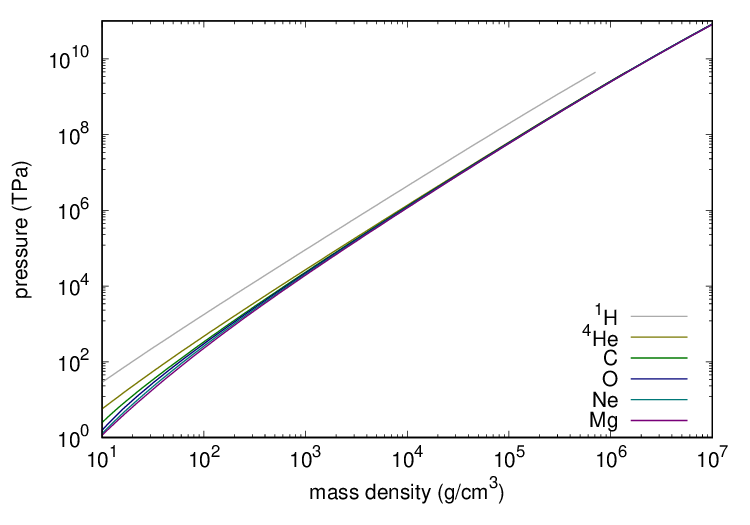}\end{center}
\caption{Comparison of atom-in-jellium cold curves calculated for 
   elements relevant to WDs.}
\label{fig:coldcmp}
\end{figure}

\begin{figure}
\begin{center}\includegraphics[scale=0.72]{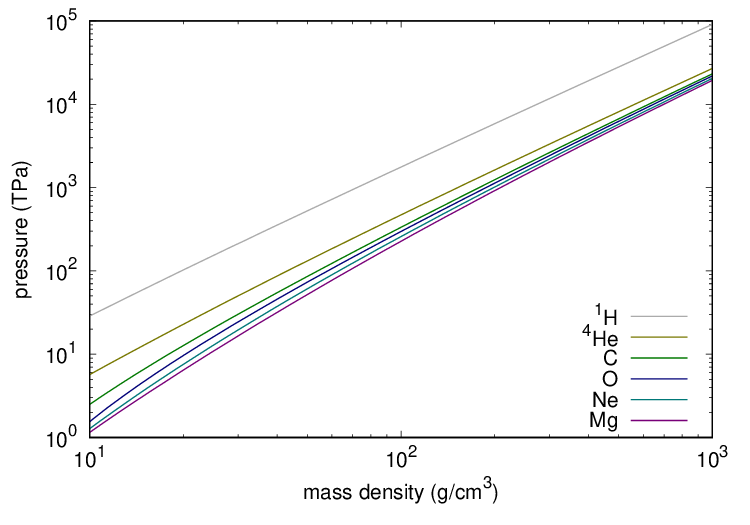}\end{center}
\caption{Comparison of atom-in-jellium cold curves calculated for 
   elements relevant to WDs
   (detail at low pressure).}
\label{fig:coldcmplo}
\end{figure}

As the program implementing the atom-in-jellium method was designed
principally for applications in warm dense matter, any numerical problems
tend to occur in cold states.
It is thus interesting to look at cold isotherms.
For all the elements considered, the isotherms were smooth over the full
compression range calculated.
The atom-in-jellium states were calculated assuming the natural atomic weight
for each element, which is close to $2uZ$ for each of the elements
apart from H ($u$).
The isotherms converged with mass density $\rho$, as one would expect
for a one-component plasma with the same $A/Z$, with that for $H$ lying
above the rest.
This convergence occurs in specific internal energy as well as pressure.
(Figs~\ref{fig:coldcmp} and \ref{fig:coldcmplo}.)

\begin{figure}
\begin{center}\includegraphics[scale=0.72]{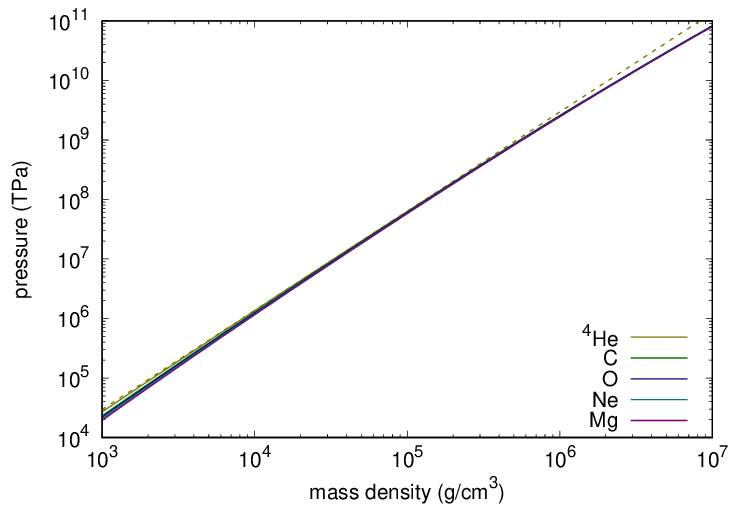}\end{center}
\caption{Comparison of atom-in-jellium cold curves calculated for 
   elements relevant to WDs
   (detail at high pressure).
   Dashed line is Thomas-Fermi-like $\rho^{5/3}$ dependence fitted to
   inflexion in $^4$He curve.}
\label{fig:coldcmphi}
\end{figure}

\begin{figure}
\begin{center}\includegraphics[scale=0.72]{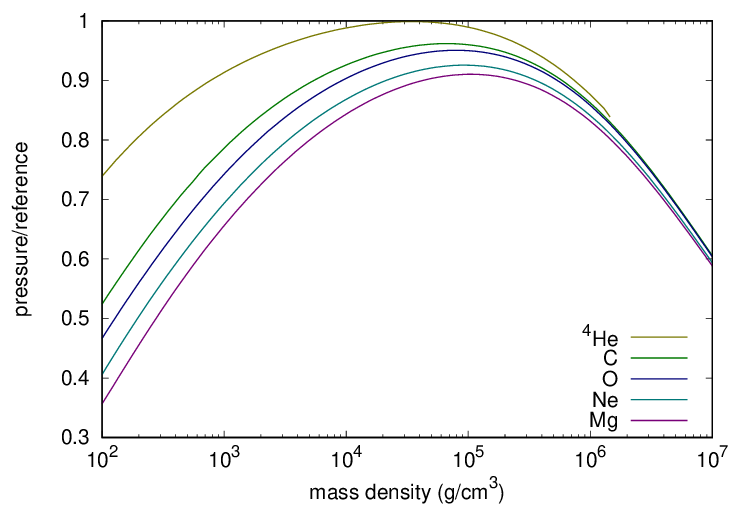}\end{center}
\caption{Atom-in-jellium cold curves for WD elements with $A=2uZ$
   scaled by $\rho^{5/3}$ fit to inflexion in $^4$He,
   showing relative softening and asymptotic behavior
   in ultrarelativistic regime.}
\label{fig:coldfcmphi}
\end{figure}

\begin{figure}
\begin{center}\includegraphics[scale=0.72]{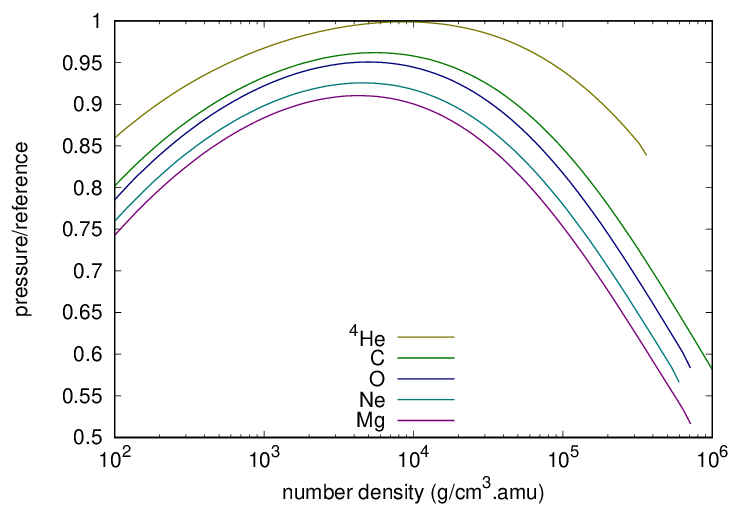}\end{center}
\caption{Previous figure, plotted against number density of ions
   instead of mass density, showing occurrence of inflexion at
   nearly-constant number density and varying dependence on $Z$.}
\label{fig:scoldfcmphi}
\end{figure}

Considered in more detail in the WD regime, the cold curves tend 
toward the $\rho^{5/3}$ dependence expected from Thomas-Fermi theory,
but then the derivative decreases and the curves asymptote toward
the $\rho^{3/2}$ dependence in the ultrarelativistic limit
\cite{Feynman1949,Carvalho2014} (Fig.~\ref{fig:coldcmphi}).
Choosing a $\rho^{5/3}$ fit to the $^4$He curve as a reference case,
although close in a logarithmic comparison, the pressure decreases
monotonically with $Z$ at constant mass density,
though in much lower proportion than the atomic number
(Fig.~\ref{fig:coldfcmphi}).
The location of the inflexion where each curve varies precisely as $\rho^{5/3}$,
but occurs close to a constant number density of ions
(Fig.~\ref{fig:scoldfcmphi}).

\section{Melt curves}
A large amount of theoretical effort has been devoted to investigating the
melt curve of matter in the one-component plasma (OCP) regime, including 
molecular dynamics \cite{Jones1996}, Monte Carlo \cite{Pollock1973},
and hypernetted chain simulations \cite{Ng1974,Kang1995},
as well as thermodynamic studies \cite{Dewitt1993,Khrapak2014}.
In the OCP, states can be analyzed in terms of the ion-ion coupling parameter
\begin{equation}
\Gamma=\frac{(Ze)^2}{\kB T a}
\end{equation}
where $a=(\frac 43 \pi n_i)^{-1/3}$,
without having to choose a specific $Z$.
Free energy calculations of the coexistence between the fluid and the solid,
predicted to be bcc, give a value of the coupling parameter for melting 
$\Gamma_m=175.0\pm 0.4$ \cite{Potekhin2000}.

\begin{figure}
\begin{center}\includegraphics[scale=0.72]{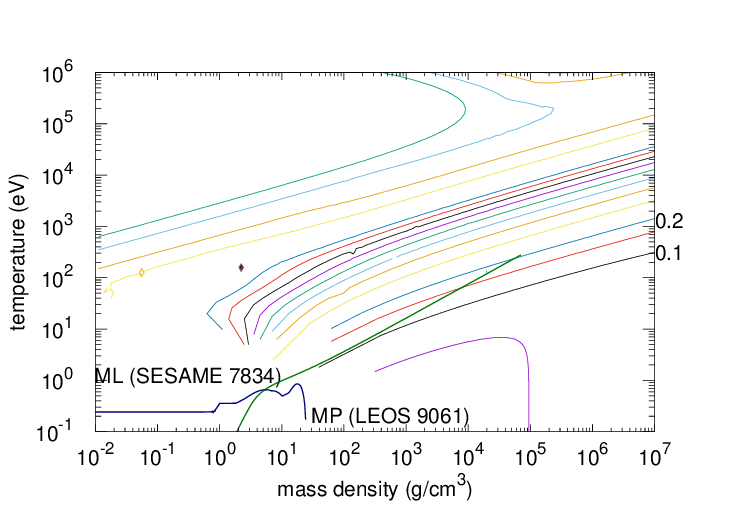}\end{center}
\caption{Contours of mean fractional ionic displacement $\bar u/\rWS$
   calculated from jellium oscillations, for carbon.
   Melt loci from previous EOS are also shown,
   based on modified Lindemann melting (ML) and a multiphase construction (MP).}
\label{fig:ccont}
\end{figure}

\begin{figure}
\begin{center}\includegraphics[scale=0.72]{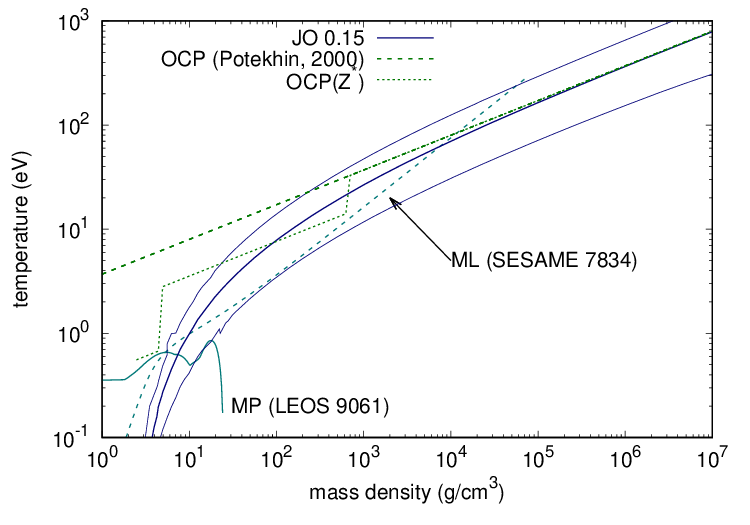}\end{center}
\caption{Melt loci for carbon:
   jellium oscillation (JO) contours of $\bar u/\rWS=0.1$, 0.2 (thin),
   and 0.15 (thick);
   OCP limit (dashed);
   OCP using ionization from atom-in-jellium (dotted); and
   loci from previous EOS,
   based on modified Lindemann melting (ML) and a multiphase construction (MP).}
\label{fig:cmelt}
\end{figure}

\begin{figure}
\begin{center}\includegraphics[scale=0.72]{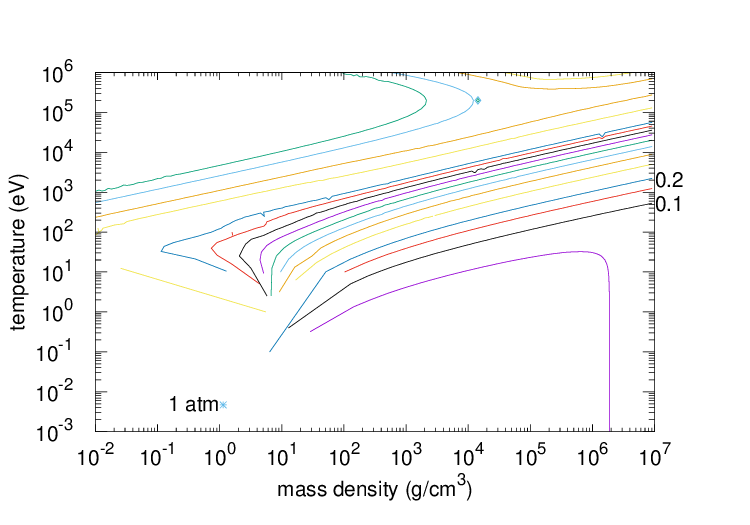}\end{center}
\caption{Contours of mean fractional ionic displacement $\bar u/\rWS$
   calculated from jellium oscillations, for oxygen, 
   showing the observed one-atmosphere melting point.}
\label{fig:ocont}
\end{figure}

\begin{figure}
\begin{center}\includegraphics[scale=0.72]{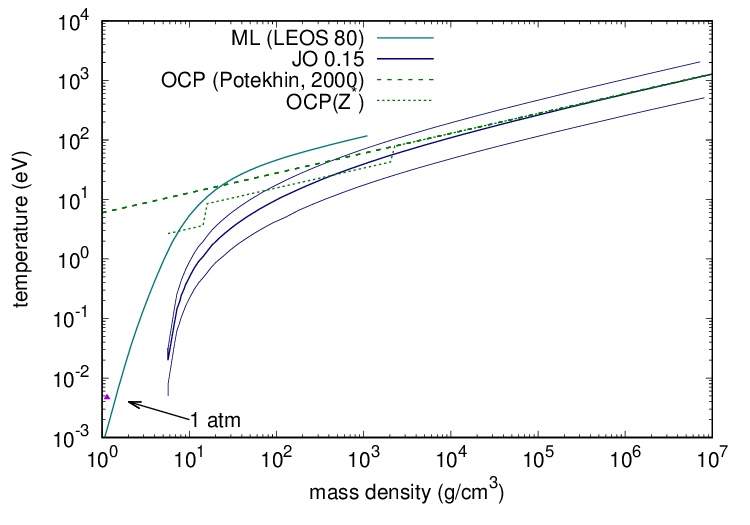}\end{center}
\caption{Melt loci for oxygen:
   jellium oscillation (JO) contours of $\bar u/\rWS=0.1$, 0.2 (thin),
   and 0.15 (thick);
   OCP limit (dashed);
   OCP using ionization from atom-in-jellium (dotted);
   a previous modified Lindemann melt curve (ML);
   and the observed one-atmosphere melting point.}
\label{fig:omelt}
\end{figure}

We have previously used a Lindemann-like criterion
applied to ion-thermal oscillations calculated using atom-in-jellium theory,
which we refer to here as jellium oscillations,
to extrapolate melt temperatures to higher compressions and pressures,
by choosing the value of mean thermal displacement of the ion $\bar u/\rWS$
to match the available data, where $\rWS$ is the Wigner-Seitz radius
\cite{Swift_ajmelt_2019}.
The ratio $\bar u/\rWS$ was found to vary between elements, with values
falling between 0.1 and 0.2
\cite{Swift_ajmelt_2019,Swift_rueos_2019,Lockard_hizeos_2019}.
For C and O at WD densities, these limiting loci bracket the OCP melt line.
For both elements, at densities above around $10^5$\,g/cm$^3$,
the OCP result (value and gradient) was reproduced with $\bar u/\rWS=0.15$.
At lower densities, where the atoms were not fully ionized,
the atom-in-jellium melt curve fell below the OCP curve, as it should to be
at all consistent with melting at more modest pressures.
Using the ionization predicted by the atom-in-jellium state instead of the
atomic number $Z$,
the OCP melting formula gave a melt locus which tracked the 
$\bar u/\rWS$ curve in an approximate, step-wise fashion over a much wider
range.
(Figs~\ref{fig:ccont} to \ref{fig:omelt}.)

Originally, the jellium oscillation melt
predictions constituted a way of extrapolating melt measurements or
predictions to higher pressure, without knowing the accuracy or where the
extrapolation was likely to break down \cite{Swift_ajmelt_2019}.
However,
the connection to OCP melting constrains the value and gradient of the melt
curve at high pressures.
If $\bar u/\rWS$ starts close to 0.15 at low pressures, as is the case for
Al \cite{Swift_ajmelt_2019}, this is likely to give a reasonable prediction
of melting up to the density limit for normal matter.
Where $\bar u/\rWS$ starts at a different value, the range 
between that value and 0.15 gives a reasonable confidence limit for the location
of the melting curve at intermediate pressures.
The possible existence of crystal structures in the solid with different free
energies means that melting may vary from the jellium oscillation prediction,
but it provides a much stronger constraint on the location than has been
available previously.
Even if this is not adequately accurate for some purposes, it provides
a good indication where more detailed theoretical or experimental studies
should be targeted.

Although not critical for WD applications,
the comparison with previous melt curves deserves some discussion.
The multiphase EOS for C \cite{Benedict2014}, {\sc leos}~9061, 
is the most rigorous general-purpose EOS available.
However, because of the computational cost of the techniques used,
the melt curve was not constructed to be physical above $\sim$20\,g/cm$^3$.
The semi-empirical, Thomas-Fermi based EOS for C, {\sc sesame}~7834,
does not include solid phase boundaries and the melt curve was
constructed using the modified Lindemann model calibrated in the region
of shock melting.
Extrapolating above $\sim$20\,g/cm$^3$,
this melt curve does not follow a contour of jellium oscillation $\bar u/\rWS$,
but it falls in the range observed for other elements.
Previous EOS for O are limited in accuracy \cite{Lockard_cryoeos_2019}.
The melt curve from the semi-empirical, Thomas-Fermi based EOS {\sc leos}~80
was constructed to pass through the observed one-atmosphere melting
temperature, but asymptotes toward a limit around twice the OCP limit
at a density two orders of magnitude below where 
the jellium oscillation calculation predicts that the limit should be reached.
This means that these EOS models are likely less suitable than those of
the present work when applied to the study of astrophysical objects
which traverse or reach, at least in part, the OCP regime.

\section{Application to white dwarfs}
EOS and solidification is of significant current interest to research on WDs.
A recent study of the population of WDs of similar mass to the sun
deduced that crystallization can be inferred as WDs cool \cite{Tremblay2019}.
Comparing simulations of the formation and cooling of WDs during the life
of the our galaxy, a feature in the luminosity population was partially explained
as a combination of the latent heat of crystallization and the release of
gravitational energy by stratification into carbon and denser oxygen.
The atom-in-jellium melt curve allows such simulations to be performed
with more rigor than the OCP melt model used previously \cite{Tremblay2019},
incorporating the effects of screening and ionization.
The atom-in-jellium isotherms are likely to be accurate over a wider range,
and suggest that the contribution to the feature in the luminosity
population from separation of species and gravitational settling is likely
to be significantly smaller than predicted in the previous study
\cite{Tremblay2019}
because of the similarity of the EOS of carbon and oxygen in this regime.
Simulations of WD cooling with simplified transport models predict a credible
luminosity feature from the latent heat but no
significant contribution from stratification
\cite{Swift2021_wdcool}.

\section{Conclusions}
In conclusion,
the atom-in-jellium method was used to construct consistent EOS 
up to the regimes of mass density and temperature occurring in WDs,
a much wider range than demonstrated previously for general-purpose EOS.
EOS were constructed for a set of elements occurring at significant levels
in WDs, and were found to asymptote to a near-unique cold curve above
$\sim 10^5$\,g/cm$^3$, for a given $A/Z$.
In particular, the isotherms of C and O were identical to within 1\%.
Our generalized Lindemann  melting criterion, based on the amplitude of jellium
oscillations, was consistent with studies of the crystallization of the OCP,
and therefore provides an estimate of the melt curve over the full range
of occurrence of normal matter.
These EOS and melt curves can be used to improve simulations of the cooling
and crystallization of WDs, and suggest that the gravitational energy
available from the separation of oxygen from carbon is significantly less than
assumed in recent work on the luminosity population distribution of WDs
of near solar mass.

\section*{Acknowledgments}
The authors would like to thank Dr~Adam Jermyn (Flatiron Institute)
and Prof.~Lars Bildsten (Kavli Institute of Theoretical Physics)
for valuable discussions.
This work was performed under the auspices of
the U.S. Department of Energy under contract DE-AC52-07NA27344.


\begin{thebibliography}{10}
\bibitem{hoteos}{For example,
   P.P.~Eggleton, Mon. Not. Roy. Astron. Soc. {\bf 163}, 279 (1973).}
\bibitem{Foley2012}{R.J.~Foley, P.J.~Challis, R.~Chornock, M.~Ganeshalingam,
   W.~Li, G.H.~Marion, N.I.~Morrell, G.~Pignata, M.D.~Stritzinger,
   J.M.~Silverman, X.~Wang, J.P.~Anderson, A.V.~Filippenko, W.L.~Freedman,
   M.~Hamuy, S.W.~Jha, R.P.~Kirshner, C.~McCully, S.E.~Persson, M.M.~Phillips,
   D.E.~Reichart, and A.M.~Soderberg,
   Astrophys. J. {\bf 767}, 1, 57 (2012).}
\bibitem{pimc}{E.L.~Pollock and D.M.~Ceperley, Phys. Rev.~B {\bf 30}, 2555 (1984).}
\bibitem{qmd}{For example,
   L.~Collins, I.~Kwon, J.~Kress, N.~Troullier, and D.~Lynch,
   Phys. Rev.~E {\bf 52}, 6202 (1995).
   }
\bibitem{Benedict2014}{L.X.~Benedict, K.P.~Driver, S.~Hamel, B.~Militzer, T.~Qi, A.A.~Correa, A.~Saul, and E.~Schwegler,
   Phys. Rev. B {\bf 89}, 224109 (2014).}
\bibitem{Driver2017}{K.P.~Driver and B.~Militzer, Phys. Rev. E {\bf 95}, 043205 (2017).}
\bibitem{Liberman1979}{D.A.~Liberman, Phys. Rev.~B {\bf 20}, 12, 4981 (1979).}
\bibitem{Swift_cowanx_2019}{D.C.~Swift, M.~Bethkenhagen, A.A.~Correa, T.~Lockard, S.~Hamel, L.X.~Benedict, P.A.~Sterne, and B.I.~Bennett,
   Phys. Rev.~E {\bf 101}, 053201 (2020).} 
\bibitem{Swift_ajeos_2019}{D.C.~Swift, T.~Lockard, M.~Bethkenhagen, R.G.~Kraus,
   L.X.~Benedict, P.~Sterne, M.~Bethkenhagen, S.~Hamel, and B.I.~Bennett,
   Phys. Rev. E {\bf 99}, 063210 (2019).}
\bibitem{sesame}{S.P.~Lyon and J.D.~Johnson, Los Alamos National Laboratory
      report LA-UR-92-3407 (1992).}
\bibitem{Lockard_cryoeos_2019}{T.~Lockard et al,
   submitted and {\tt arXiv:1906.09516}}
\bibitem{leos}{D.A.~Young and E.M.~Corey, J.~Appl. Phys. {\bf 78}, 3748 (1995).}
\bibitem{Feynman1949}{R.P.~Feynman, N.~Metropolis, and E.~Teller,
   Phys. Rev. {\bf 75}, 10, 1561-1573 (1949).}
\bibitem{Carvalho2014}{S.M.~de~Carvalho, M.~Rotondo, J.A.~Rueda, and R.~Ruffini,
   Phys. Rev. C {\bf 89}, 015801 (2014).}
\bibitem{Jones1996}{M.D.~Jones and D.M.~Ceperley, Phys. Rev. Lett. {bf 76}, 24, 4572-4575 (1996).}
\bibitem{Pollock1973}{E.L.~Pollock and J.P.~Hansen, Phys. Rev. {\bf A8}, 3110 (1973).}
\bibitem{Ng1974}{K.-C.~Ng, J.~Chem. Phys. {\bf 61}, 7, 2680-2689 (1974).}
\bibitem{Kang1995}{H.-S.~Kang, J.~Chem. Phys. {\bf 103}, 9370 (1995).}
\bibitem{Dewitt1993}{H.E.~Dewitt, Contrib. Plasma Phys. {\bf 33}, 5-6, 399-408 (1993).}
\bibitem{Khrapak2014}{S.A.~Khrapak and A.G.~Khrapak, Phys. Plasmas {\bf 21}, 104505 (2014).}
\bibitem{Potekhin2000}{A.Y.~Potekhin and G.~Chabrier,
   Phys. Rev.~E {\bf 62}, 6, 8554 (2000).}
\bibitem{Swift_ajmelt_2019}{D.C.~Swift, T.~Lockard, R.F.~Smith, C.J.~Wu, and L.X.~Benedict,
   Phys. Rev. Research {\bf 2}, 023034 (2020).}
\bibitem{Swift_rueos_2019}{D.C.~Swift et al, {\tt arXiv:1909.05391}}
\bibitem{Lockard_hizeos_2019}{T.~Lockard et al, in preparation.}
\bibitem{Tremblay2019}{P.-E.~Tremblay, G.~Fontaine, N.P.G.~Fusillo, B.H.~Dunlap, B.T.~G\"ansicke, M.A.~Hollands, J.J.~Hermes, T.R.~Marsh, E.~Cukanovaite, and T.~Cunningham,
   Nature {\bf 565}, 202 (2019).}
\bibitem{Swift2021_wdcool}{D.C.~Swift et al, in preparation.}
\end{thebibliography}
\end{document}